\documentclass[journal = nalefd, layout = traditional]{achemso} 

\setkeys{acs}{usetitle=true}
\setkeys{acs}{maxauthors=20}

\usepackage{graphicx}
\usepackage{hyperref}
\usepackage{soul}
\usepackage{amsmath}
\usepackage[usenames,dvipsnames]{color}
\usepackage{amssymb}
\usepackage{xcolor}
\usepackage{units}

\newcommand{\ket}[1]{|#1\rangle}

\title{Observation of Coexistence of Yu-Shiba-Rusinov States and Spin-Flip Excitations}

\author{Shawulienu Kezilebieke}
\affiliation{Department of Applied Physics, Aalto University School of Science, 00076 Aalto, Finland}

\author{Rok \v{Z}itko}
\affiliation{Jo\v{z}ef Stefan Institute, Jamova 39, SI-1001 Ljubljana, Slovenia}
\alsoaffiliation{Faculty  of Mathematics and Physics, University of Ljubljana,
Jadranska 19, SI-1000 Ljubljana, Slovenia}

\author{Marc Dvorak}
\affiliation{Department of Applied Physics, Aalto University School of Science, 00076 Aalto, Finland}

\author{Teemu Ojanen}
\affiliation{Department of Applied Physics, Aalto University School of Science, 00076 Aalto, Finland}
\alsoaffiliation{Computational Physics Laboratory, Physics Unit, Faculty of Engineering and Natural Sciences, Tampere University, P.O. Box 692, FI-33014 Tampere, Finland}

\author{Peter Liljeroth}
\email{Email: peter.liljeroth@aalto.fi}
\affiliation{Department of Applied Physics, Aalto University School of Science, 00076 Aalto, Finland}


\begin{document}


\begin{abstract}
We investigate the spectral evolution in different metal phthalocyanine molecules on NbSe$_2$ surface using scanning tunnelling microscopy (STM) as a function of the coupling with the substrate. For manganese phthalocyanine (MnPc), we demonstrate a smooth spectral crossover from Yu-Shiba-Rusinov (YSR) bound states to spin-flip excitations. This has not been observed previously and it is in contrast to simple theoretical expectations. We corroborate the experimental findings using numerical renormalization group calculations. Our results provide fundamental new insight on the behavior of atomic scale magnetic/SC hybrid systems, which is important for e.g. engineered topological superconductors and spin logic devices.
\end{abstract}
\maketitle

Precise control of the properties of magnetic impurities on surfaces, such as the spin state and magnetic anisotropy, is one of the ultimate goals in fabricating atomic or molecular scale devices for data storage or computing purposes. However, the properties of magnetic impurities are strongly influenced by the atomic environment. In the extreme case, the interactions with the environment (substrate) can create entirely new electronic states, such as the Kondo effect \cite{Schneider1998kondo,Crommie1998kondo,Franke2011}, or the formation of Yu-Shiba-Rusinov (YSR) bound states on superconductors \cite{YU1965,Shiba1968,Rusinov1969,Yazdani1997,Heinrich2018review}. YSR states have received intense interest as it has become possible to create artificial designer structures, where the interaction between the YSR states gives rise to Majorana modes \cite{Nadj-Perge2014,Ruby2015_prl_2,Pawlak2016_Majorana,Kezilebieke2018,Ruby2017NanoLett,Menard2017,Ruby2018_PRL,Wiesendanger2018_Majorana}.
The YSR states are very sensitive to the immediate environment of the impurity spin and give information on the role of the local environment on the exchange interaction $J$ of an impurity spin with a superconductor \cite{Yazdani1997,Ji2008,Franke2011,Heinrich2013,Menard2015,Ruby2015_PRL,Ruby2016_PRL,Cornils2017_PRL,vanderZant2017PRL,Choi2017_NatCommun,Choi2018_PRL,Ruby2018_PRL,Kezilebieke2018,Wiesendanger2018_Majorana,Senkpiel2018arXiv,Etzkorn2018arXiv,Liebhaber2019arxiv,Schneider2019arXiv}. The bulk of recent experimental work on YSR states on superconducting (SC) substrates has demonstrated that the strength of the exchange interaction $J$ can be significantly influenced by a small change in the adsorption site of the  impurity or by spacers between the  impurity and  substrate \cite{Franke2011,Hatter2015,Hatter2017,Heinrich2013,Heinrich2015,Choi2018_PRL,Farinacci2018_PRL,Loth2018_NanoLett,Senkpiel2018arXiv,Etzkorn2018arXiv,Liebhaber2019arxiv,Schneider2019arXiv}.

\begin{figure}[!t]
	\centering
	\includegraphics{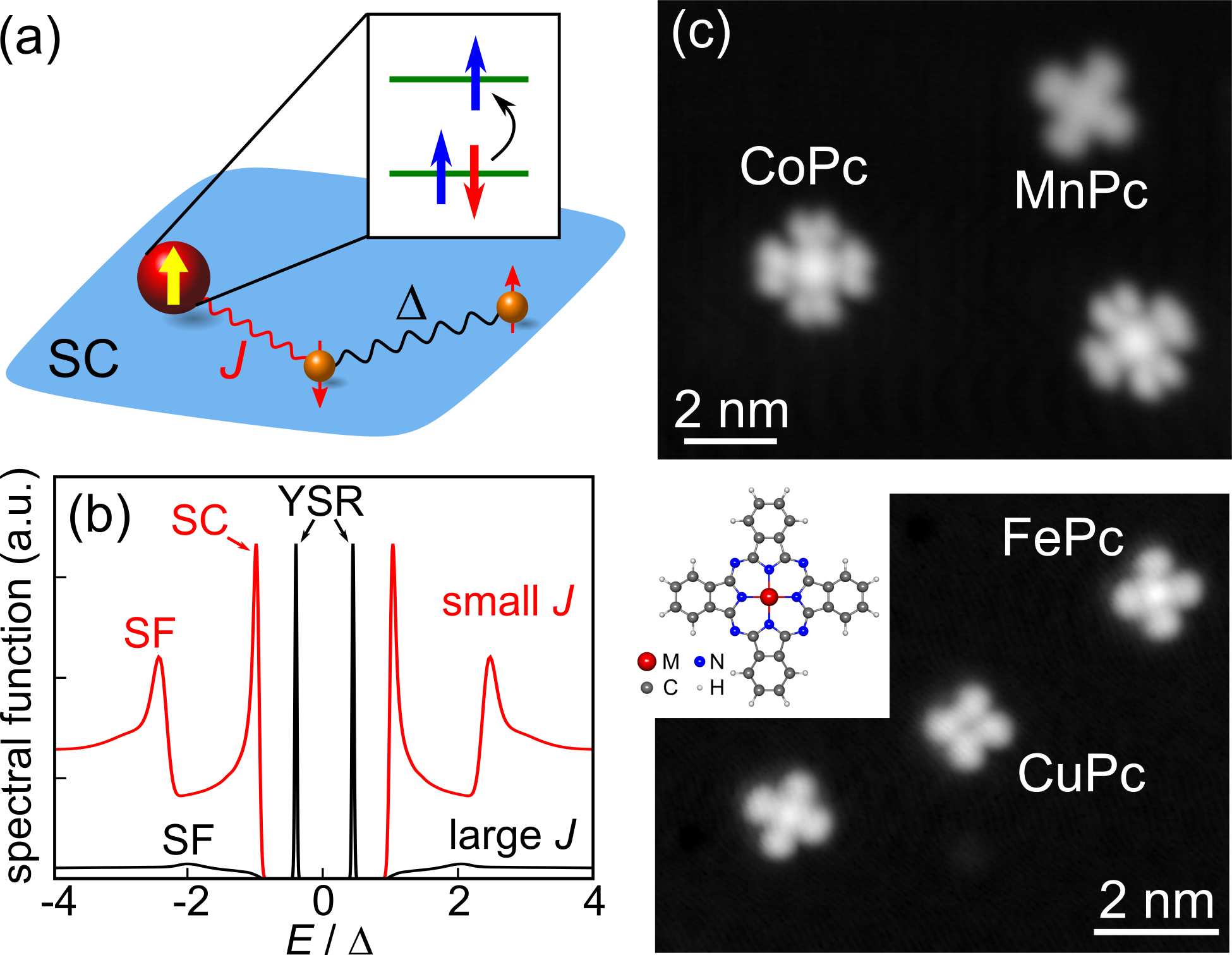}
	\caption{(a) Impurity spin is exchange coupled (strength $J$) to the superconducting substrate  (gap $\Delta$). (b) Impurity induces pairs of bound states symmetric with respect to $E_F$ within the gap of the quasiparticle excitation spectrum (black curve, states marked with YSR). Even at vanishing $J$, for impurities with $S\geq 1$, the internal spin degrees of freedom can result in symmetric features with respect to $E_F$ outside the SC gap (red curve, spin-flip (SF) excitations). (c) Topographic STM images of the molecules used in this study (imaging set-point 0.5 V / 2 pA). Inset: Schematic of a MPc molecule.}
	\label{fig1}
\end{figure}

Figure~\ref{fig1}a illustrates how the exchange coupling $J$ with the substrate competes with the superconducting (SC) pairing energy $\Delta$. The interaction of the local spin with the Cooper pairs gives rise to a low-lying excited state within the gap of the quasiparticle excitation spectrum \cite{YU1965,Shiba1968,Rusinov1969,Yazdani1997,Heinrich2018review}. When $J$ is decreased, the direct interaction of the local spin with the Cooper pairs is reduced and the YSR states merge with the SC coherence peaks. In the simplified theory by Yu, Shiba and Rusinov, the position of the YSR state is $E_\mathrm{YSR}=\Delta (1-\alpha^2)/(1+\alpha^2)$ with $\alpha$ proportional to $J$, $\alpha=\pi \rho J S/2$, where $\rho$ is the normal-state density of states of the substrate at the Fermi level and $S$ is the impurity spin. The bound state results from the spin-dependent scattering of Bogoliubov quasiparticles on the impurity and is thus associated with the longitudinal part of the exchange interaction, $J S_z s_z$, where $s$ represents the spin-density of the substrate electrons at the impurity position. Furthermore, internal spin transitions in combination with magnetic anisotropy can give rise to symmetric features with respect to $E_F$ outside the superconducting gap (Fig.~\ref{fig1}b) \cite{Heinrich2004,Hirjibehedin2006,Hirjibehedin2007,Heinrich2013,Heinrich2015,wiesendanger2009_review,Ternes2015,Ternes2017_review}. These are associated with the spin-flip (SF) events, whereby the spin projection changes by $\pm 1$ \cite{Heinrich2004,Ternes2015,Ternes2017_review}. The renormalization of the magnetic anisotropy, associated with the transverse part of the exchange interaction $J (S^+ s^- + S^- s^+)$,
approximately follows a $D_\mathrm{eff}=D_0 [1-\beta (\rho J)^2 + \ldots]$ dependence. 
In the simple picture, the relative magnitudes of these two channels (YSR and spin-flip) are not constrained and, in principle, both of these effects should be observed simultaneously \cite{Berggren2014,Fransson2015_theory}. While both YSR states and spin-flip excitations have been observed on the same experimental system, where the exchange coupling is changed by the adsorption site of the magnetic impurity,\cite{Cornils2017_PRL} they have not been observed simultaneously in a single configuration. This suggests that the complete picture of the interplay between these two effects is more complicated, requiring a full many-body treatment of the quantum mechanical spin degree of freedom interacting with a (gapped) continuum of electrons. 

Here, we experimentally demonstrate smooth cross-over from YSR states to spin-flip excitations in metal phthalocyanine molecules (MPcs) on NbSe$_2$ surface. Using scanning tunneling microscopy (STM), we tune the exchange coupling strength $J$ and follow the spectral evolution from the YSR states to intrinsic quantum spin states characterized by well-developed spin-flip excitations. Our results provide detailed understanding of the low-energy quantum states in magnetic/SC hybrid systems and could have significant ramifications for the design and control of atomic-scale magnetic devices.

\begin{figure}[!t]
	\centering
	\includegraphics[width=.7\textwidth]{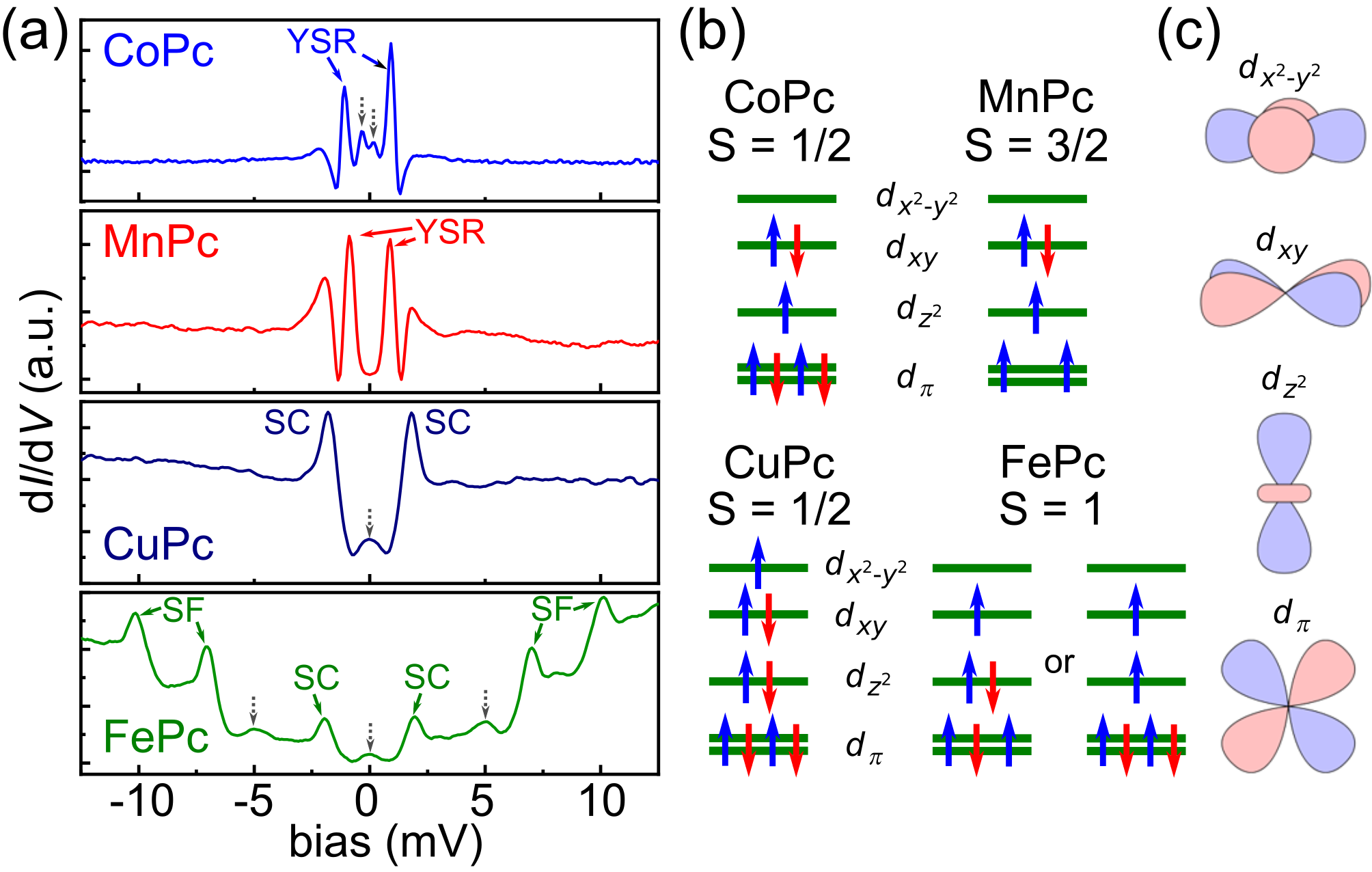}
	\caption{(a) Differential conductance spectra (d$I$/d$V$) recorded over the center of the different MPcs with a SC tip. The YSR states and spin-flip (SF) excitations are marked. Gray dotted arrows mark transitions due thermal excitation of carriers across the SC gap \cite{Franke2011}. (b) The ground state spin configurations of the different MPcs. For FePc, we show two possible spin configurations (having similar energy). (c) Illustration of the different $d$-orbital symmetries.}
	\label{fig2}
\end{figure}

Figure \ref{fig1}c shows topographic STM images of isolated MPc molecules on NbSe$_2$, see Supporting Information (SI) for details. Their topographic appearance already reflects differences that allow us to classify them into two groups: The metal ion appears as a protrusion in FePc, CoPc and MnPc and as a depression in CuPc depending on the coupling of the out of plane $d$- orbitals with the tip states \cite{Lu1996,Kuegel2015_PRB}. We characterize the spin states of different MPcs by recording differential conductance spectra (d$I$/d$V$ curves) with a SC tip (Fig.~\ref{fig2}a). For CoPc and MnPc, there are two peaks at symmetric bias voltages within the SC gap. These sub-gap peaks are due to the formation of YSR states and indicate a sizable magnetic interaction with the SC substrate caused by an unpaired spin in the $d_{z^2}$-orbital (see Fig.~\ref{fig2}b for the spin states of the molecules). This orbital is subject to strong coupling with the electronic states of the substrate due to its symmetry, while the spins on the $d_{xy}$- and $d_{x^2-y^2}$-orbitals are expected to be only weakly coupled \cite{Kuegel2014,Kuegel2015_PRB}. The d$I$/d$V$ curve taken on CuPc shows an unperturbed SC gap of NbSe$_2$ due to the absence of unpaired spin in the $d_{z^2}$ orbital (Fig.~\ref{fig2}b). As a $S=1/2$ system, CuPc is also not expected to show any spin-flip excitations. 

While the d$I$/d$V$ curves taken on FePc do not show YSR states, there are remarkable features outside the SC gap. Their symmetric appearance points to inelastic excitations \cite{Jaklevic1966,Stipe1998,Heinrich2004}. Interestingly, unlike in the previous studies that required decoupling the magnetic molecules using an extra organic ligand as a spacer \cite{Heinrich2013,Heinrich2015}, we observe these signals already when the molecule is directly adsorbed on the SC substrate. Since FePc is also expected to have a dominant $d_{z^2}$-character, the molecule-substrate interaction should be similar to CoPc and MnPc. The surprising absence of YSR states on FePc suggests that the magnetic interaction with the SC substrate is actually weak, which would be at odds with an unpaired spin occupying a $d_{z^2}$-orbital. FePc has spin triplet $S=1$ electronic ground state and DFT calculations suggest that FePc on NbSe$_2$ has the same spin as in the gas phase (SI for details). Two different spin-configurations separated by 80 meV have been proposed as the ground state \cite{Fernandez-Rodriguez2015,Tsukahara2016} (Fig.~\ref{fig2}b). The absence of YSR states is consistent with the predicted lower energy configuration with two electrons on the $d_{z^2}$-orbital \cite{Fernandez-Rodriguez2015}. This ground state spin configuration can be altered by slight differences in the molecular ligand field or the interaction with the substrate as shown by YSR states observed on a related iron porphyrin molecule on Pb(111) substrate.\cite{Farinacci2018_PRL}

We have verified that the observed transitions result from inelastic spin excitations by acquiring the d$I$/d$V$ spectra under an external magnetic field ($B$) perpendicular to the sample surface (see SI for details). The $B$-field dependence of the energies of the first and second feature is consistent with $S=1$ with transverse anisotropy. Fitting the data with a phenomenological spin Hamiltonian $H_\mathrm{eff}=g\mu_B B S_\gamma+DS_z^2+E(S_x^2-S_y^2)$, where $g$ is the Land\'{e} $g$ factor, $\mu_B$ the Bohr magneton and $S_\gamma$ the spin component along the field direction, we obtain $D=5.5$ meV and $E=1.4$ meV indicating easy-plane magnetic anisotropy. The positive value of $D$ is comparable to the bulk value (in contrast to the measurements on oxidized Cu(110) surface giving $D<0$)\cite{Tsukahara2009,Tsukahara2016}. 
      
The exchange coupling between the magnetic impurity and the substrate can be modulated by changing the adsorption site of the molecule. \cite{Franke2011,Franke2013PRB,Hatter2015,Hatter2017,Heinrich2013,Heinrich2015,Kuegel2018_npj} We successfully positioned CoPc and FePc molecules on different adsorption sites through STM manipulation (SI Fig.~S3), which affects the energy positions of both the YSR states and the inelastic features. On CoPC, the YSR states can even change between particle- or hole-like character depending on the adsorption site, similarly to the reported results on MnPc adsorbed on Pb(111) \cite{Hatter2015}. On FePc, adsorption site causes variations of $E$ and $D$ with typical values in the range of $1.2-2.7$ meV and $4.2-9.0$ meV, respectively.   

\begin{figure}[!t]
\centering
\includegraphics[width=.8\textwidth]{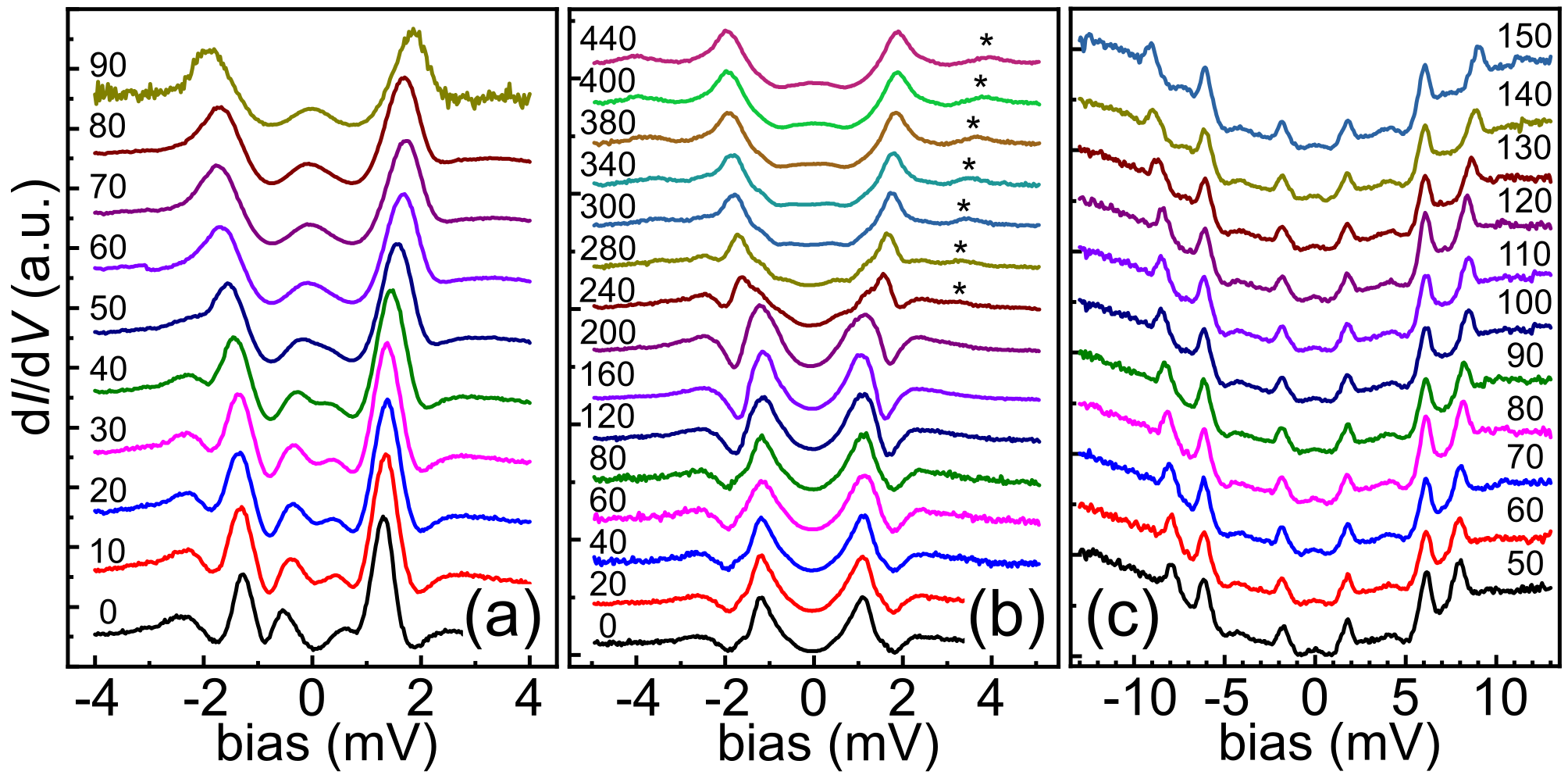}
\caption{Normalized d$I$/d$V$ spectra recorded over CoPc (set point: V = 20 mV, I = 300 pA) (a), MnPc (set point: V =20 mV, I = 200 pA) (b), and FePc (set point: V = 50 mV, I = 200pA) (c) at different tip-sample distances from far (bottom) to close (top). Initial tip-sample distance given by the set-point conditions, then the tip is approached by a distance of $z_\mathrm{offset}$ indicated in the figure (values in pm).}
\label{fig3}
\end{figure}

Due to the sensitivity of the YSR states and spin-excitations, they can be tuned continuously by the force exerted by the STM tip \cite{Ternes2011,Farinacci2018_PRL,Loth2018_NanoLett}. We expect to have attractive forces between the tip and the molecule;\cite{Gross2011_Science,Boneschanscher2012_ACSNano} this would result in pulling the molecule away from the substrate and reduction of the exchange coupling between the molecule and the substrate upon decreasing tip-molecule distance. Fig.~\ref{fig3} shows a series of tunneling spectra measured at different tip-sample distances above the central ion of the MPc molecule. The d$I$/d$V$ spectra on a CoPc molecule, Fig.~\ref{fig3}a, show a shift of the YSR resonances towards higher bias. At $z_\mathrm{offset} \approx 60$ pm, the YSR resonances have merged with the SC coherence peaks at the gap edge. The d$I$/d$V$ spectra on MnPc (Fig.~\ref{fig3}b) also show a clear shift of the YSR resonance towards the SC gap edge. At $z_\mathrm{offset} \approx 240$ pm, as the YSR resonance is merging with the SC coherence peak, a new symmetric pair of peaks is emerging outside the gap (marked with $*$). Finally, on FePc (Fig.~\ref{fig3}c), the two spin excitation energies monotonously increase with decreasing tip-sample distance. These variations in the YSR and spin excitations states of the MPc molecules are caused by the interaction of the STM tip causing the metal ion to be pulled towards the STM tip \cite{Ternes2011,Farinacci2018_PRL,Loth2018_NanoLett}. This has the largest effect on the out-of-plane $d$-orbitals ($d_{xz/yz}$ and $d_{z^2}$), where the overlap with the substrate wave function will be strongly affected.  

With CoPc ($S=1/2$), the interaction with the tip reduces the coupling between the $d_{z^2}$-orbital and the NbSe$_2$ and the YSR states move to the SC coherence peaks ($J$ decreases) (Fig.~\ref{fig3}a). With FePc ($S=1$), in addition to modifying the exchange coupling $J$, the tip-sample interaction might also affect the energies of the $d$-orbitals slightly. This causes the bare magnetic anisotropy $D_0$, which is an admixture of low-lying excited states of the molecules \cite{Wang1993}, to be directly affected by the tip-sample distance. In this system, it is not possible to distinguish whether the effective anisotropy $D_\mathrm{eff}$ changes dominantly through the variation of $D_0$ or $J$.

With MnPc ($S=3/2$), the coupling strength $J$ decreases as the tip-sample distance decreases leading to the migration of the YSR states towards the SC gap edge and the recovery of the SC coherence peaks. When the YSR peaks are close to the gap edges, the spectra also show new symmetric features outside the gap, which are due to spin-flip excitations. In MnPc, at zero external magnetic field, the axial magnetic anisotropy splits the spin states $M_s=\pm1/2$ and $M_s=\pm 3/2$ by an energy separation of $2D$, which corresponds to the observed spin excitation. This gives $D \approx 0.7$ meV, which is close to the bulk value \cite{Barraclough1974} (more detailed analysis below). There could also be inelastic excitations in this energy range corresponding to molecular vibrations (phonons). We can rule this out by experiments under an external magnetic field, which also indicate a positive value of $D>0$ (see SI for details).

Conventional picture of the YSR states and spin-flip excitations neglecting quantum fluctuations does not predict a cross-over behaviour from the YSR states to spin-excitations. These conduction channels should be independent and readily observable at the same time. To resolve this disagreement with our experimental results, we describe the magnetic impurity by a multi-orbital Anderson model for the relevant $d$-shell orbitals \cite{Georges2013_review}. This can be reduced in a given charge state to an effective model that takes the form of a Kondo Hamiltonian with the spin degree of freedom, $S$, only \cite{Zitko2016_PRB,Zitko2017_PRB} (see SI for details). Since the total spin operator has contributions from all $d$-orbitals, it is in general exchange coupled to different symmetry-adapted combinations of states from the substrate with different values of Kondo coupling strength, $J_i$. These differences are due to the unequal orbital energies and hybridization strengths. Only those $J_i$ that are sufficiently large need to be retained; in the problems considered here, a single orbital is always strongly dominant, as evidenced by the presence of a single pair of sub-gap YSR peaks. We also take into account the spin-orbit coupling that leads through orbital excitations to residual magnetic-anisotropy terms. We solve the resulting single-channel anisotropic Kondo model with high spin $S$ using the numerical renormalization group (NRG) method giving a numerically exact solution \cite{wilson1975,bulla2008,hofstetter2000}.

\begin{figure}[!t]
	\centering
	\includegraphics{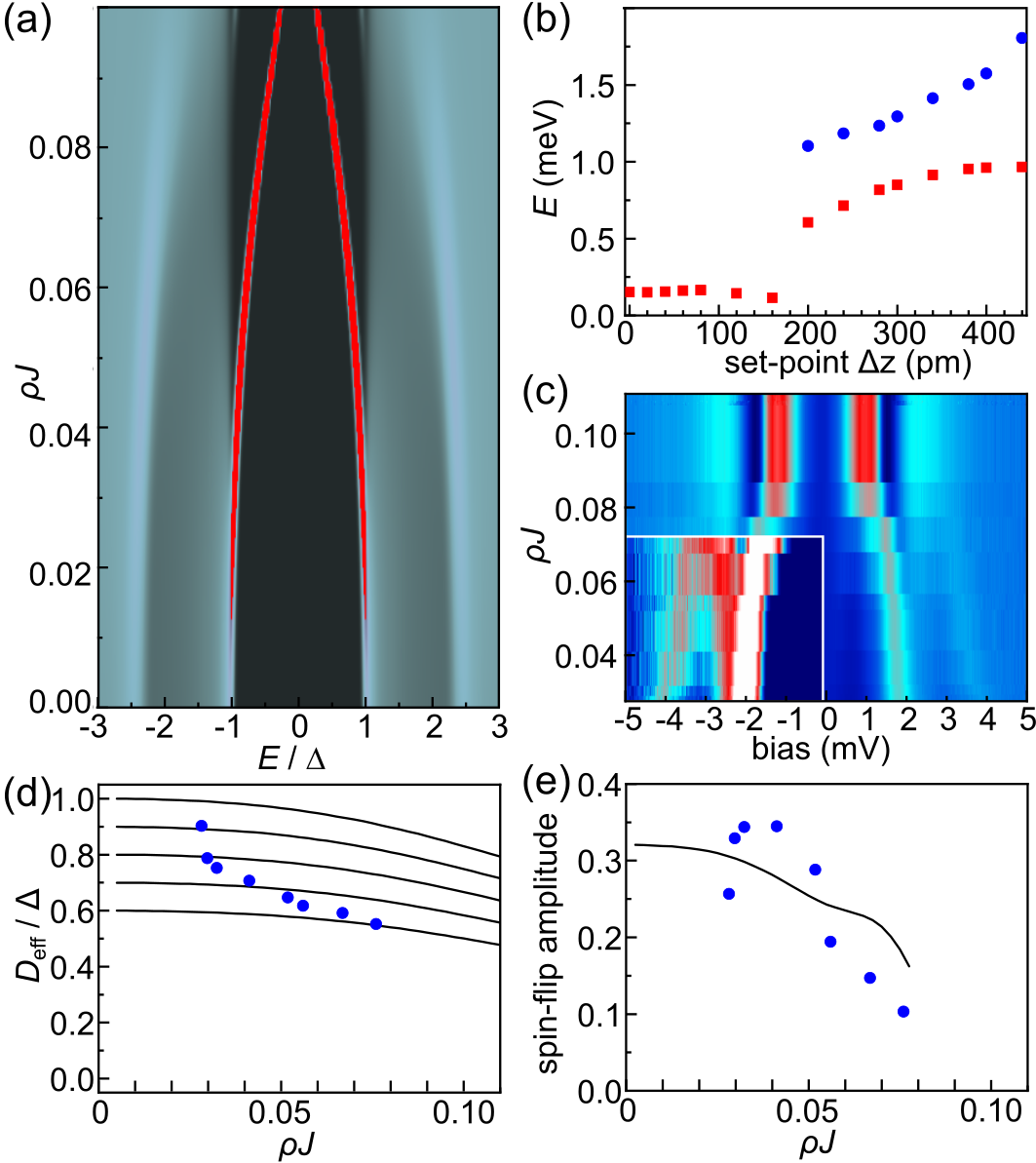}
	\caption{(a) Theoretical spectral functions for a magnetic impurity ($S=3/2$) on a superconductor with varying exchange coupling $J$ (spectra normalized by the value at high energy). (b) Extracted energies of the YSR resonances (red squares) and the spin-flip excitations (blue circles) from the experiments shown in Fig.~\ref{fig3}b. (c) Colour-scale plot of the measured tunneling conductance as a function of the exchange coupling $\rho J$ estimated from the energies of the YSR resonances (area inside the white rectangle is shown in enhanced contrast). (d) Comparison between the experimental $D_\mathrm{eff}$ (symbols) and the calculated values (solid lines with $D_0/\Delta=0.6-1.0$) scaled by the SC gap $\Delta$ as a function of the exchange coupling with the substrate. (e) Extracted experimental values (blue circles) and calculation results (black line) of the amplitude of the spin-flip transition as a function of the exchange coupling with the substrate.}
	\label{fig4}
\end{figure}
Figure~\ref{fig4}a shows the calculated spectral functions as a function of the exchange coupling for a magnetic impurity with $S=3/2$ (\emph{e.g.} MnPc). In the absence of anisotropy ($D=E=0$), the binding of a Bogoliubov quasiparticle would lead to an emergence of a YSR bound state with the spin reduced by $1/2$ (YSR screening from $S=3/2$ to $S=1$), giving rise to a single pair of YSR peaks. The phenomenology in the isotropic case is similar to that in the classical model, where the impurity is described as a static local magnetic field that binds Bogoliubov quasiparticles of the opposite spin direction. The many-body character of the sub-gap states is, however, revealed in the presence of magnetic anisotropy, which leads to the splitting of the sub-gap $S=1$ multiplet into the low-energy $\ket{S_z=0}$ state and the high-energy $(1/2)(\ket{S_z=1}\pm\ket{S_z=-1})$ states \cite{Zitko2011_PRB}.
For $D \ll \Delta$, such splitting is directly observable \cite{Hatter2015}. When the anisotropy is large, as is the case here, the high-energy states are pushed instead into the continuum and only the $\ket{S_z=0}$ sub-gap state is observable. Furthermore, one finds additional features outside the gap that are related to the transitions within the unscreened $S=3/2$ multiplet \cite{Heinrich2013}. These spectral steps correspond to the spin-flip excitations, similar to the those in systems with normal-state substrates \cite{Hirjibehedin2007,Otte2008_NatPhys}. The exchange coupling to the substrate determines the energy shift as well as the lifetime broadening of the excited states \cite{Heinrich2013,Oberg2013}. We model NbSe$_2$ as a soft-gap superconductor with a small but finite concentration of sub-gap states, hence the excited spin states have even for $\omega_{sf}=2D<2\Delta$ relatively short lifetimes  compared to hard-gap superconductors, because the dominant decay channel (emission of particle-hole excitations) is here open. For this reason, a finite value of $J$ leads to significant broadening of the spin-flip excitations. It is difficult to resolve both spectral features, considering the required magnitude of $J$ for the YSR states to be visibly separated from the gap edges. 

The theory curves in Fig.~\ref{fig4}a can be compared with the experiments shown in Fig.~\ref{fig3}b. The energies of the YSR peaks and spin-flip excitations extracted from the experimental results are shown in Fig.~\ref{fig4}b (the SC gap of the tip has been subtracted). At small values of $z_\mathrm{offset}< 200$ pm, only the YSR resonances are visible. As the tip is approached further, the YSR peaks merge with SC gap edges and spin-flip features emerge at a bias voltage between 3 and 4 mV. We can estimate the exchange coupling with the substrate by comparing the theoretical and experimental YSR energies (SI Fig.~S8). This allows us to convert set-point $\Delta z$ to effective exchange coupling at that tip-molecule separation. The experimental data is replotted as a function of $\rho J$ in Fig.~\ref{fig4}c, which can be directly compared with Fig.~\ref{fig4}a (note that the bias axis in Fig.~\ref{fig4}c still contains the off-set due to the superconducting gap of the tip).

We can also plot the renormalized (effective) magnetic anisotropy $D_\mathrm{eff}$ and the intensity of the spin-flip transitions as a function of the estimated $\rho J$ (Figs.~\ref{fig4}d,e). This is possible as the energy of the YSR state depends only very weakly on the precise value of $D$ in the experimentally relevant range (SI Fig.~S8). The comparison with theoretical results (solid lines in Fig.~\ref{fig4}d) yields a value of $D_0/\Delta \sim 0.7$ for the non-renormalized (bare) value of the magnetic anisotropy. The experimental values actually deviate from the theoretical trend expected for a fixed value of $D_0$. This is the case for all couplings $J$, but more particularly for small $J$, which corresponds to the largest tip-molecule interaction in the experiment. It is likely that the interaction with the tip distorts the molecular geometry or induces charge transfer with the metal ion resulting in a change in $D_0$. The presence of the YSR resonances thus allows us to disentangle the two contributions to the variation of $D_\mathrm{eff}$. The extracted value $D_0$ as a function of the tip-sample distance is plotted in the SI Fig.~S9.

The intensity of the spin-flip transition (Fig.~\ref{fig4}e) shows a monotonous decrease with increasing $J$ (except the first two points, where the reason is again likely to be the tip-molecule interaction). The value at low $J$ is close to the theoretically expected value of 0.4 for pure spin-flip excitations on a normal metal substrate (see SM Fig.~S10 for details) \cite{Ternes2015,Ternes2017_review}. Comparing the experimental values with the theoretical predictions again highlights the strong correspondence between theory and experiments. 

In conclusion, we have demonstrated co-existence and a smooth evolution from the YSR states to spin-flip transitions in MnPc molecules as the coupling with the NbSe$_2$ substrate is reduced. The excitation energies reveal a significant renormalization of the anisotropy by the exchange coupling and it has a strong effect on the excited spin lifetime. The spin-flip excitations are broadened and washed out at higher values of $J$ making simultaneous detection with the YSR states difficult. Our results provide fundamental new insight on the behavior of atomic scale magnetic/SC hybrid systems. This is important for e.g. the design of engineered topological superconductors consisting of magnetic atoms on SC substrates \cite{Nadj-Perge2014,Ojanen2015_PRL,Wiesendanger2018_Majorana} and achieving long spin life- and coherence times in spin logic devices \cite{Baumann2015_Science,Paul2016_NatPhys,Natterer2017_Nature}.

\begin{suppinfo}
Experimental and computational methods, and additional results.
\end{suppinfo}

\begin{acknowledgement}
	This research made use of the Aalto Nanomicroscopy Center (Aalto NMC) facilities and was supported by the European Research Council (ERC-2017-AdG no.~788185 ``Artificial Designer Materials''), Academy of Finland (Academy professor funding nos.~318995 and 320555, Academy Research Fellow no.~256818 and Postdoctoral Researcher nos.~309975 and 316347), and the Aalto University Centre for Quantum Engineering (Aalto CQE). Our DFT calculations were performed using computer resources within the Aalto University School of Science “Science-IT” project and the Finnish CSC-IT Center for Science. R.\v{Z}. acknowledges the support of the Slovenian Research Agency (ARRS) under P1-0044 and J1-7259.
	
\end{acknowledgement}

\bibliography{shiba}

\end{document}


\newpage	

	\subsection{Additional experimental details}
	Sample preparation and subsequent STM experiments were carried out in an ultrahigh vacuum system with a base pressure of $\sim$10$^{-10}$ mbar. The $2H$--NbSe$_2$ single crystal (HQ Graphene, the Netherlands) was cleaved in situ by attaching a tape to the crystal surface and pulling the tape in vacuum in the load-lock chamber using the sample manipulator. MPc molecules (Sigma-Aldrich) were deposited from an effusion cell held at 390$^{\circ}$C onto a freshly cleaved NbSe$_2$ at room temperature.  
	
	After the MPc deposition, the sample was inserted into the low-temperature STM (Unisoku USM-1300) and all subsequent experiments were performed at $T=4.2$ K. STM images were taken in the constant-current mode. d$I$/d$V$ spectra were recorded by standard lock-in detection while sweeping the sample bias in an open feedback loop configuration, with a peak-to-peak bias modulation of $50-100$ $\mu$V at a frequency of 709 Hz. The procedure for acquiring a spectrum was as follows: the tip was moved over the molecule at the imaging parameters (e.g. $V = 0.6$ V and $I = 5$ pA), the tip-sample distance was reduced by changing the setpoint to e.g. 200 pA at 100 mV. Finally, after disconnecting the feedback at the beginning the d$I$/d$V$ spectrum, the tip-sample distance was decreased by a further $50 - 100$ pm ($z_\mathrm{offset}$) to increase the signal to noise ratio. The detailed numbers are given in the figure captions.
	
	The NbSe$_2$ tip was prepared by indenting the tip into the NbSe$_2$ surface by a few nanometers while applying a voltage of 10~V. Manipulation of the MPc was carried out by placing the tip above the centre of the molecule with a bias voltage of 0.1 V and the current was increased to 1 nA with the feedback engaged. The tip was then dragged towards the desired location.
	
	\newpage
	\subsection*{Geometries and spin densities from DFT calculations}
	Density functional theory calculations were performed with the FHI-AIMS computational package \cite{Blum2009,Havu2009} and the PBE generalized gradient approximation for the exchange-correlation functional \cite{Perdew1996}. We used the standard "light" numerical settings and basis sets of numeric atomic-centered orbitals tested and recommended by FHI-AIMS. Periodic NbSe$_2$ supercells were sampled with a $2\times2$ $k$-point grid centred on the $\Gamma$ point. Van der Waals interactions were included by the post-SCF Tkatchenko-Scheffler correction \cite{Tkatchenko2009}. Before computing the electronic structure, all atomic forces were relaxed to $<0.01$ eV/{\AA}.
	
	\begin{figure}[!h]
		\centering
		\includegraphics [width=.95\textwidth]{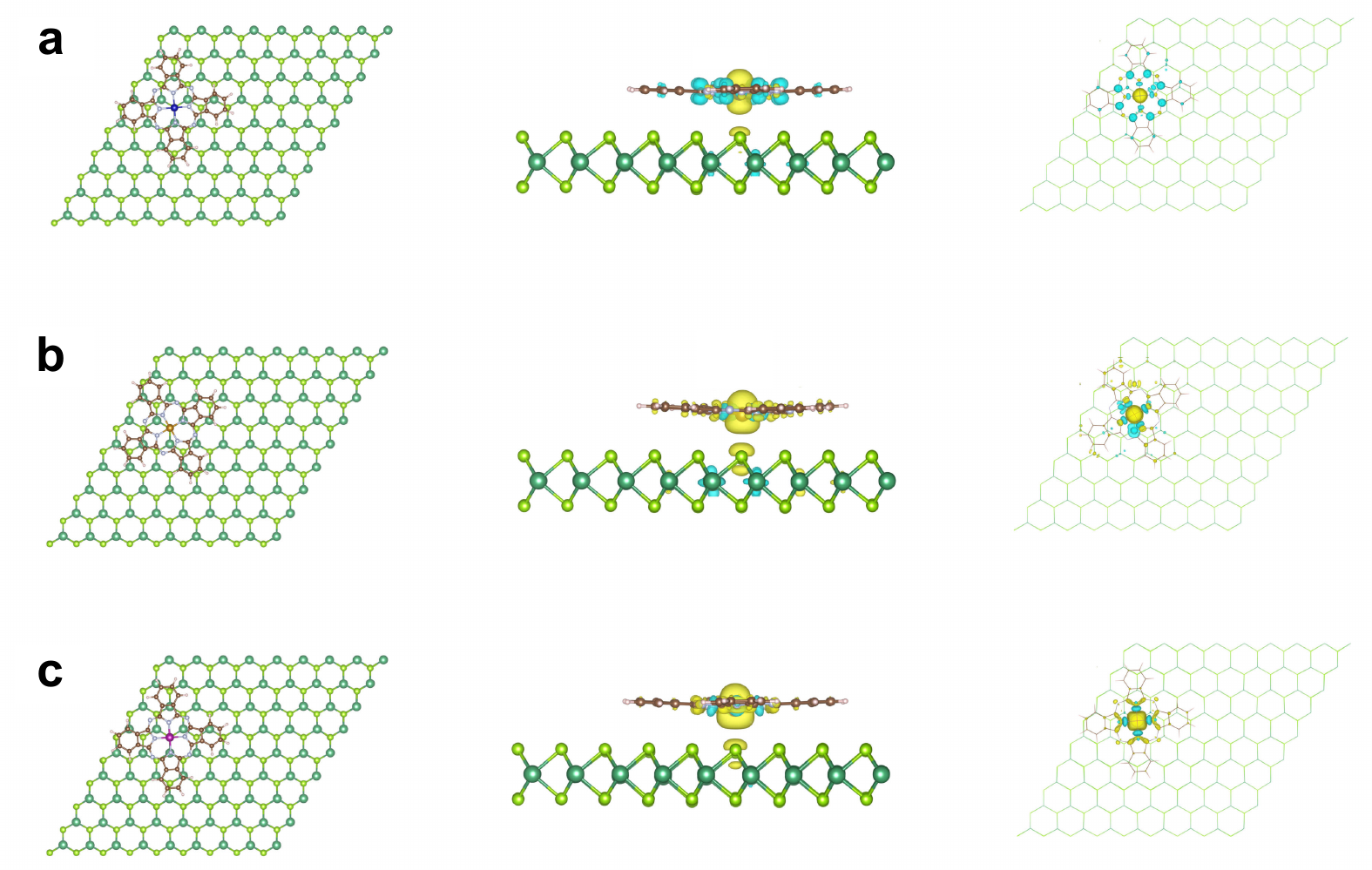}
		\caption{(a,b,c) Results for CoPc (a), FePc (b), and MnPc (c) computed with density functional theory (DFT) \cite{Blum2009,Havu2009}. The total computed spins for the systems are $S=0.41$, $S=1.22$, and $S=1.72$, respectively. The localized metal atom-projected (Co, Fe, or Mn) spins are $S = 0.49$, $S=1.10$, and $S=1.70$, respectively. The localized atom-projected spins differ by $<0.1$ compared to their gas phase counterparts with the same DFT parameters and numerical settings.}
		\label{si_dft}
	\end{figure}

	\subsection*{Experiments on FePc in external magnetic field}
	
	\begin{figure}[!b]
		\centering
		\includegraphics [width=0.9\textwidth] {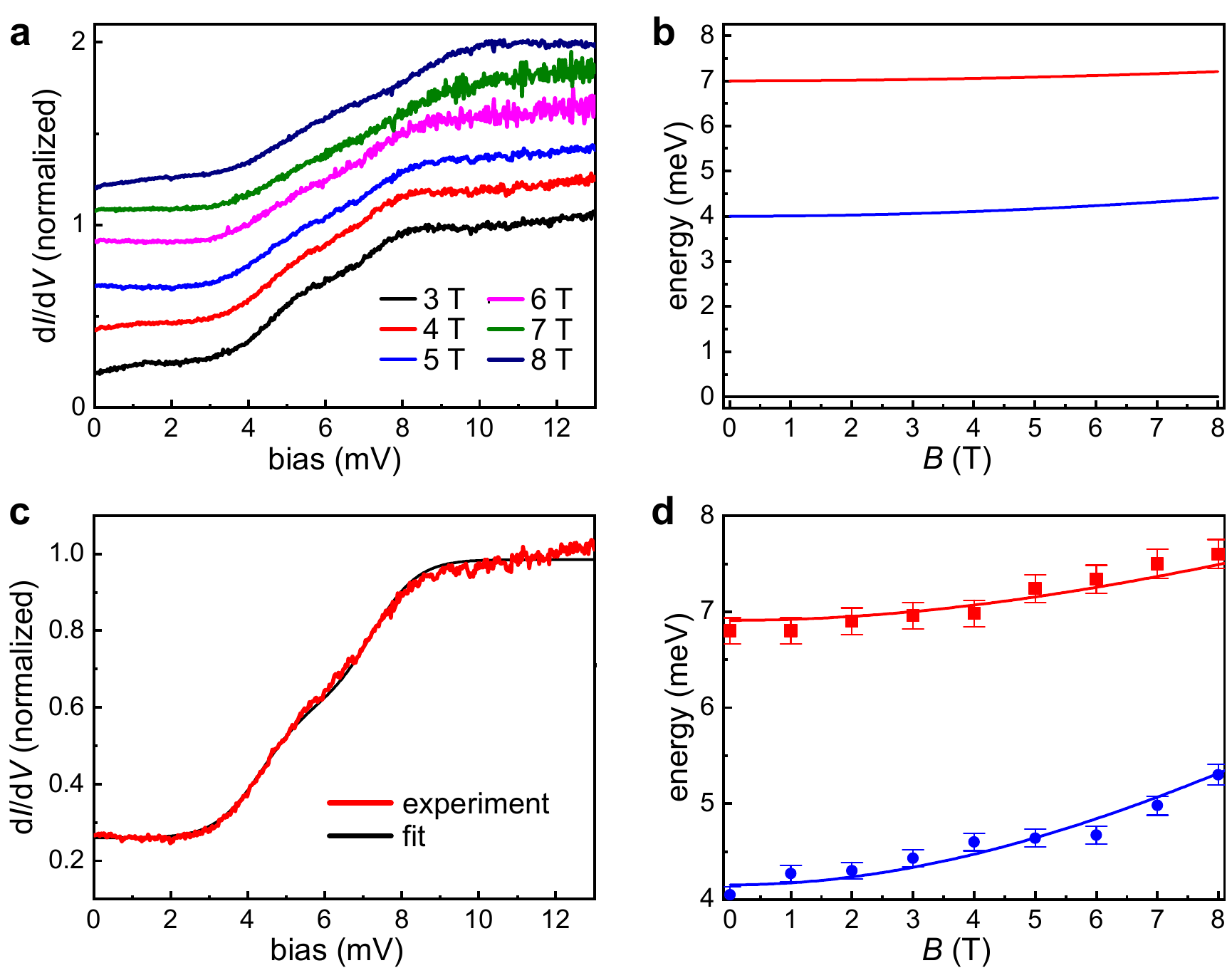}
		\caption{Spin-flip excitations in FePc under external magnetic field. (a) d$I$/d$V$ spectra of FePc taken as a function of B$_z$ (3 - 8 T, perpendicular to the sample surface). Each spectrum was acquired directly above the central Fe atom of the FePc. The spectra are vertically offset for clarity (set-point conditions: 20 mV / 1 nA). (b) Schematic level diagram for an $S=1$ system with $g=2$, $D= 5$ meV and $E= 1$ meV. Due to the selection rule $m_S$ =$\pm 1$, there are two allowed transitions from the ground state. (c) d$I$/d$V$ spectrum measured under an external field of 5 T showing the splitting of the inelastic spin excitation due to the presence of transverse magnetic anisotropy term $E$. The black line represents the fit of the spectra \cite{Ternes2015}. (d) Spin-flip energies as a function of the external magnetic field. From the fit, we obtain $g=3.6$, $D=5.5$ meV, and $E=1.4$ meV.}
		\label{FigSI1}
	\end{figure}
	
	We can verify that the observed transitions in FePc do indeed correspond to inelastic spin excitations by acquiring d$I$/d$V$ spectra at different magnetic ($B$) field strengths aligned perpendicular to the sample surface. The energies of the first and second feature change with the external field, demonstrating that these are spin excitations (see  Fig.~\ref{FigSI1}). 
	The zero-field splittings (ZFS) are well described by the following spin Hamiltonian: $$H_\mathrm{eff}=DS_z^2+E(S_x^2-S_y^2)$$ where $D$ is the axial ZFS constant to determine the magnetic anisotropy, $E$ is the transverse anisotropy, and $S_z, S_x, S_y$ are the spin operators. The selection rule for the spin excitations implies that they can only occur between the states differing by $m_S$ =$\pm 1$. For $S=1$ spin-state, we would expect only one step in the absence of transverse anisotropy ($E=0$) and at $B=0$T. Transverse anisotropy ($E>0$) mixes the states and therefore a second step can occur. Hence, the d$I$/d$V$ spectra on FePc on NbSe$_2$ suggest spin-state $S=1$ with transverse anisotropy. We fitted our data  using a phenomenological spin Hamiltonian \cite{Heinrich2004,Hirjibehedin2007,Ternes2015,Ternes2017_review}
	$$H_\mathrm{eff}=g\mu_B B S_\gamma+DS_z^2+E(S_x^2-S_y^2)$$ where $g$ is the Landg\'{e} $g$ factor, $\mu_B$ the Bohr magneton, and $S_\gamma$ is the component of spin along the direction of the magnetic field. Using the numerical code by M. Ternes \cite{Ternes2015}, we obtain $D=5.5$ meV, and $E=1.4$ meV as the best fit with the experimental data. The positive value of $D$ indicates easy plane magnetic anisotropy. 
	
	\newpage
	\subsection*{Effect of the adsorption site on the YSR states and the spin-flip excitations}
	
	\begin{figure}[!h]
		\centering
		\includegraphics [width=0.9\textwidth] {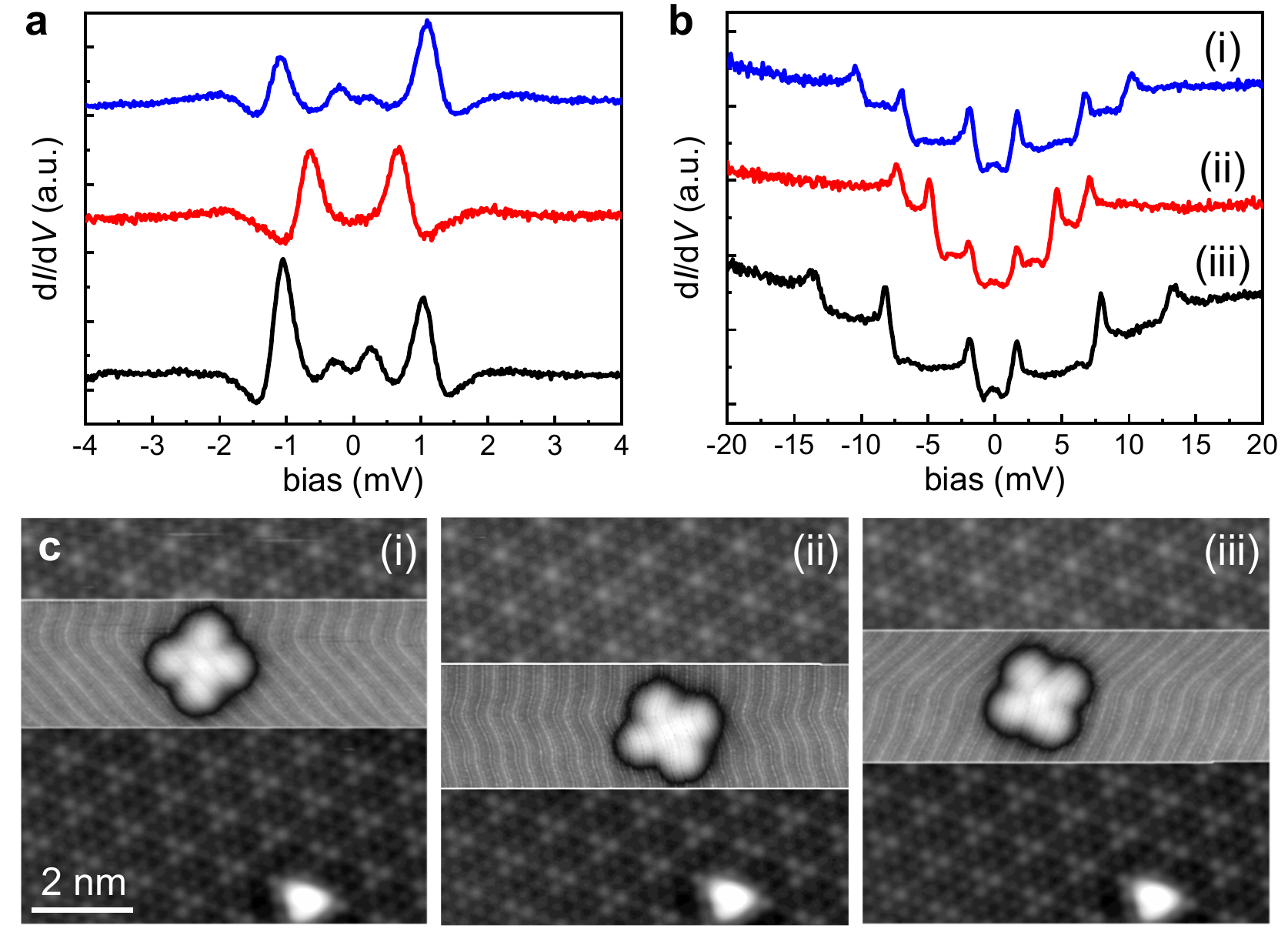}
		\caption{Effect of the adsorption site on the YSR states and the spin-flip excitations. (a) d$I$/d$V$ spectra recorded on the same CoPc molecule at different adsorption sites (feedback opened at 10 mV / 1 nA; the spectra are shifted for clarity). All spectra show YSR states inside the superconducting gap. The YSR states change their excitation character between particle- or hole-like depending on the adsorption site of the CoPc molecule on NbSe$_2$, indicating different signs of the potential scattering term. (b) d$I$/d$V$ spectra taken on the same FePc molecule at different adsorption sites (feedback opened at 20 mV / 1 nA; the spectra are shifted for clarity). None of the spectra show any sharp features within the superconducting gap. However, adsorption site causes variations of both $E$ and $D$ with typical values in the range of $1.2-2.7$ meV and $4.2-9.0$ meV, respectively. (c) Determination of the adsorption site corresponding to the three spectra shown in panel b on FePc.}
		\label{FigSI2}
	\end{figure}
	
	\newpage

	\subsection*{Experiments on MnPc in external magnetic field}
	
	\begin{figure}[!b]
		\centering
		\includegraphics [width=0.95\textwidth] {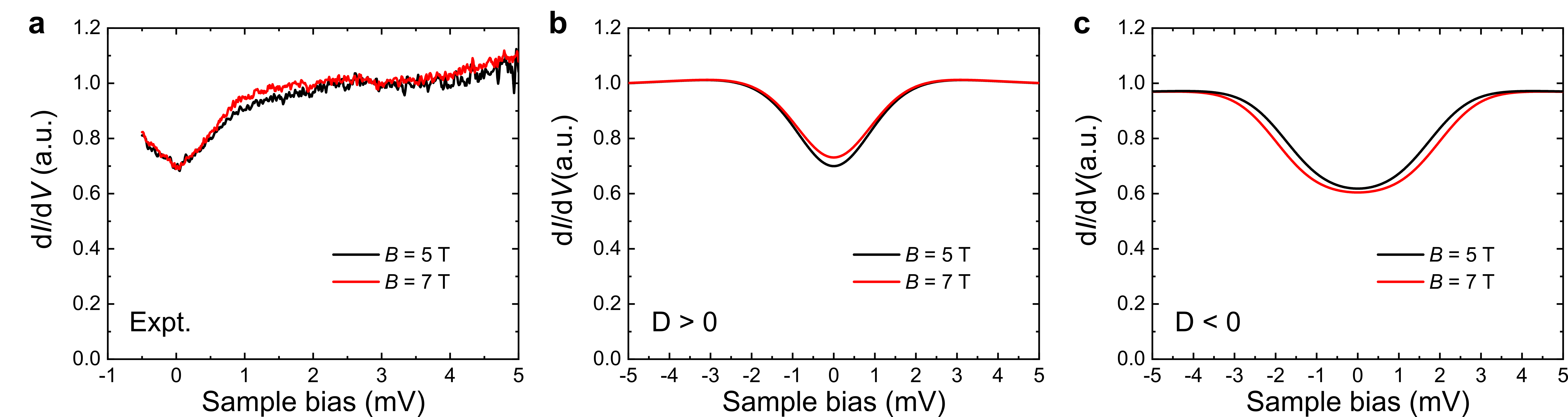}
		\caption{Spin-flip excitations in MnPc under external magnetic field. (a) Experimental results on a MnPc molecule on NbSe$_2$ substrate under external magnetic field showing the spin-excitations. (b,c) Simulated spectra using $D = 0.6$ meV (b) and $D = -0.6$ meV (c) ($g=2$, $T = 4.2$ K). }
		\label{FigSI-MnPc}
	\end{figure}
	
	It is conceivable that there would be inelastic excitations in the energy corresponding to the features seen outside the superconducting gap on MnPc that correspond to molecular vibrations (phonons). To support our assignment of this feature to a magnetic excitation, we have carried out experiments under an external magnetic field. These experiments are difficult as they require 1) close tip-sample distances to move into the regime where we observe the spin-flip transitions and 2) we lose the enhanced energy resolution due to the superconducting tip (as superconductivity is quenched by the external field). Fig.~\ref{FigSI-MnPc}a are examples of the spectra on MnPc at external fields of 5 T and 7 T (where both the substrate and the tip are normal metals). Fig.~\ref{FigSI-MnPc}b,c also shows simulated spectra for $S=3/2$ with $D = 0.6$ meV (panel b) illustrating reasonable agreement with the experiments (considering the experimental difficulties). The spin-excitation feature is weakly dependent on the magnetic field and shift towards zero bias with increasing external field strength. This also confirms that $D > 0$ (we would expect opposite and stronger magnetic field dependence for $D < 0$ as shown in Fig.~\ref{FigSI-MnPc}c). The simulations were carried out using a phenomenological spin Hamiltonian \cite{Heinrich2004,Hirjibehedin2007,Ternes2015,Ternes2017_review} and using the numerical code by M. Ternes \cite{Ternes2015}.

	\subsection*{Numerical renormalization group calculations}
	
	The impurity model is solved using the numerical
	renormalization group (NRG) method \cite{wilson1975,bulla2008}. The
	NRG is a numerical procedure for solving quantum impurity models that
	is based on a logarithmic discretization of the continuum of electrons
	(here itinerant electrons from the substrate hybridized with the
	molecular states). The
	discretization is controlled by the parameter $\Lambda>1$ that
	determines the coarseness of the frequency grid, $\omega_n \sim
	\Lambda^{-n}$. The discretized Hamiltonian is transformed to a linear
	tight-binding representation that is diagonalized iteratively. One can
	compute static properties (expectation values of various operators),
	thermodynamics, as well as dynamic properties (spectral functions). The
	results, in particular spectra, can be significantly improved by
	performing several calculations for interleaved discretizations grids
	and averaging the results. This tends to remove the discretization
	artifacts which are periodic functions in $\ln{\omega}$ with period
	$\sqrt{\Lambda}$; using $N_z=2^k$ grids leads to a good cancellation
	of the fundamental frequency and the first $k-1$ harmonics.
	
	\newcommand{\mD}{\mathcal{D}}
	
	In this work, the calculations have been performed with the ``NRG
	Ljubljana'' implementation of the technique. We used $\Lambda=2$,
	$N_z=16$, and kept up to 12000 states/multiplets (or states with
	energies up to $10\omega_N$, with $\omega_N$ the characteristic energy
	scale at the $N$-th step of the iteration). The conduction band was
	assumed to have a constant density of states $\rho=1/(2\mD)$ ranging
	in energy from $-\mD$ to $\mD$; $\mD$ also serves as the energy unit.
	The gap is fixed at $\Delta=10^{-3}\mD$ unless otherwise specified.
	The spectra were computed using the density-matrix NRG algorithm
	\cite{hofstetter2000}.
	
\begin{figure}[!h]
		\centering
		\includegraphics [width=0.6\textwidth] {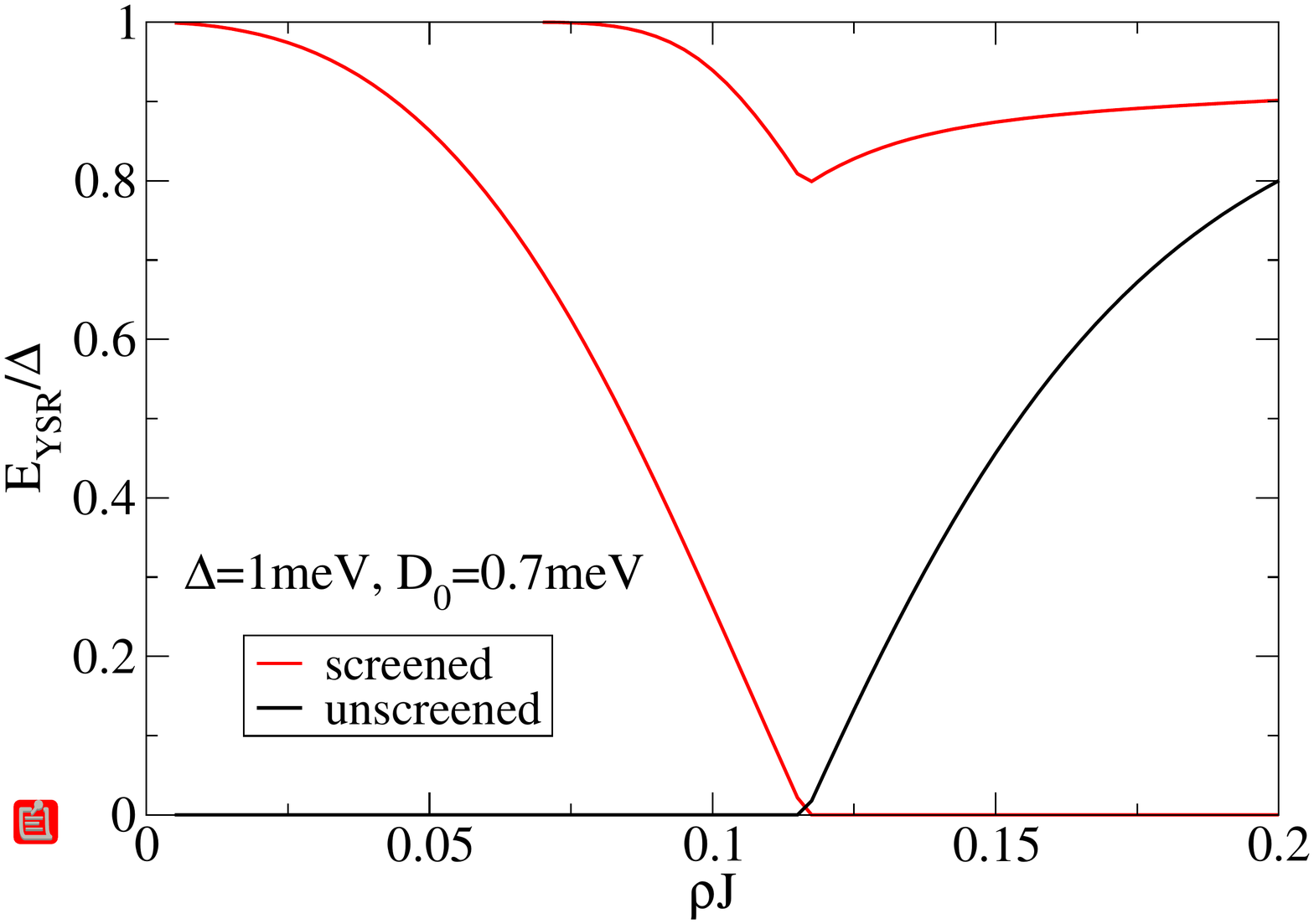}
		\caption{Energies of the many-body states, referenced to the ground-state
			energy of the system.
			This diagram shows the energies of all states below the edge of the
			continuum excitations. The lowest state is hence by definition the
			ground state, while the excited states give rise to the YSR peaks in
			the spectra. Black lines correspond to the states resulting from the
			$S=3/2$ multiplet after anisotropy splitting (i.e., the $S_z = \pm
			1/2$ pair). Red lines correspond to the states resulting from the
			impurity state ``screened'' by binding one Bogoliubov quasiparticle
			from the continuum. These states originate from the $S=1$ multiplet
			and are separated into a $S_z=0$ (lower energy) and a $S_z=\pm 1$
			(higher energy) subsets. The $S_z=0$ YSR state is present for all $J$,
			the $S_z=\pm 1$ pair starts emerging from the continuum at $\rho J
			\approx 0.08$. At $\rho J\approx 0.12$, a quantum phase transition
			between the $S_z = \pm 1/2$ and $S_z=0$ many-body states occurs.
		}
		\label{FigSIRZ4}
	\end{figure}

	The energies of the sub-gap states can be directly extracted from the
	renormalization-group flow diagrams in the NRG. This approach is
	significantly more accurate than determining the peak positions in the
	spectra, and does not suffer from any broadening artifacts. Example results are shown in Fig.~\ref{FigSIRZ4}
	
\begin{figure}[!h]
		\centering
		\includegraphics [width=0.6\textwidth] {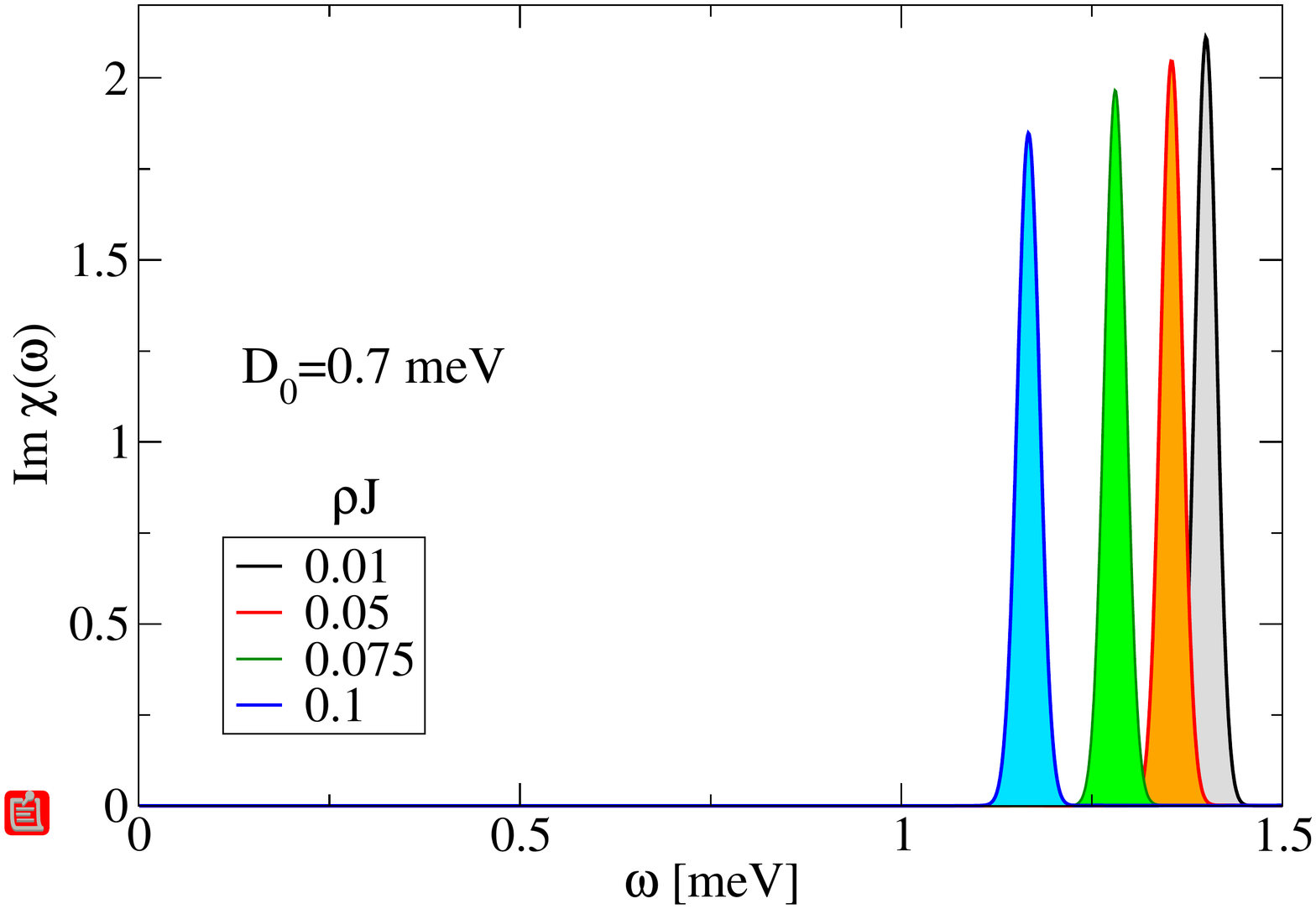}
		\caption{Dynamical spin susceptibility for a range of exchange couplings $J$.
			Here we plot the imaginary part of the transverse component of the
			dynamical spin susceptibility which quantifies the spin-flip
			transitions of the system. The peak position corresponds to
			$\omega_\text{sf}=2D_\mathrm{eff}$, i.e., the renormalized spin-flip
			energy. The peak-width is related to the life-time of the spin
			excitations, but here it is overbroadened for numerical reasons.}
		\label{FigSIRZ5}
	\end{figure}

	The positions of the spin-flip excitation energies can be determined
	from the spectral function, for instance by locating the peak position
	or the inflection point that approximately corresponds to the center
	of the step. Due to strongly changing line-shape of the equilibrium
	spectral function, this procedure is somewhat ill-defined,
	and furthermore suffers from possible broadening artifacts. In
	theoretical calculation using the NRG one can, however, use a more
	robust and direct approach for very accurately extracting the energies
	of spin excitations. It consists of computing the transverse part of
	the dynamical spin susceptibility function
	\begin{equation}
		\chi_\perp(\omega) = \langle\langle S^x ; S^x \rangle\rangle_\omega.
	\end{equation}
	The spin excitation energies can then be directly read off from the
	peak position, which is unique and very well defined. This is illustrated in Fig.~\ref{FigSIRZ5}
	\newpage
	
	\newpage
	
	\subsection*{Properties of the anisotropic $S=3/2$ Kondo model}
\begin{figure}[!h]
		\centering
		\includegraphics [width=0.8\textwidth] {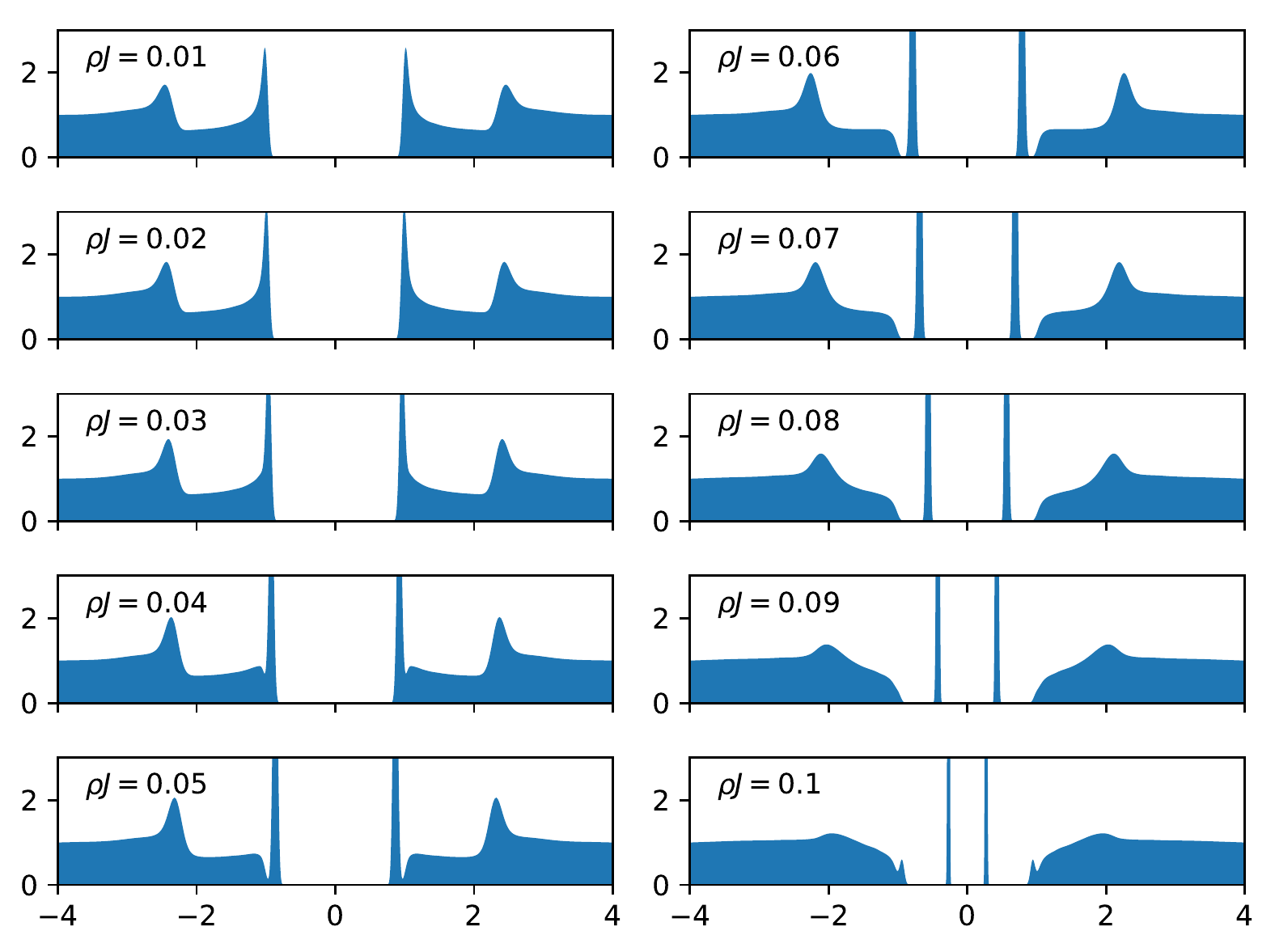}
		\caption{Spectral functions for a range of exchange coupling strengths.
			These plots correspond to horizontal line-cuts of the density plot in
			Fig.~4a of the main text. They show more clearly how the YSR state
			detaches from the band edge of the continuum of Bogoliubov
			excitations, and how the spin-flip line-shape evolves (in particular
			for $\rho J \geq 0.06$, i.e., after the emergence of YSR as a
			well-defined sub-gap state). At $\rho J \sim 0.1$, a new sub-gap
			feature starts to detach from the band edge: this is an additional YSR
			state (of type $|S_z=1\rangle \pm |S_z=-1\rangle$) that results from the 
			magnetic anisotropy splitting of the $S=1$ YSR multiplet. This parameter range
			is, however, not experimentally relevant.
		}
		\label{FigSIRZ3}
	\end{figure}

	The spectral functions for a range of values of the exchange coupling for $S=3/2$ are shown in Fig.~\ref{FigSIRZ3} (with parameter values corresponding to the experiments on MnPc).
	
\begin{figure}[!h]
		\centering
		\includegraphics [width=0.6\textwidth] {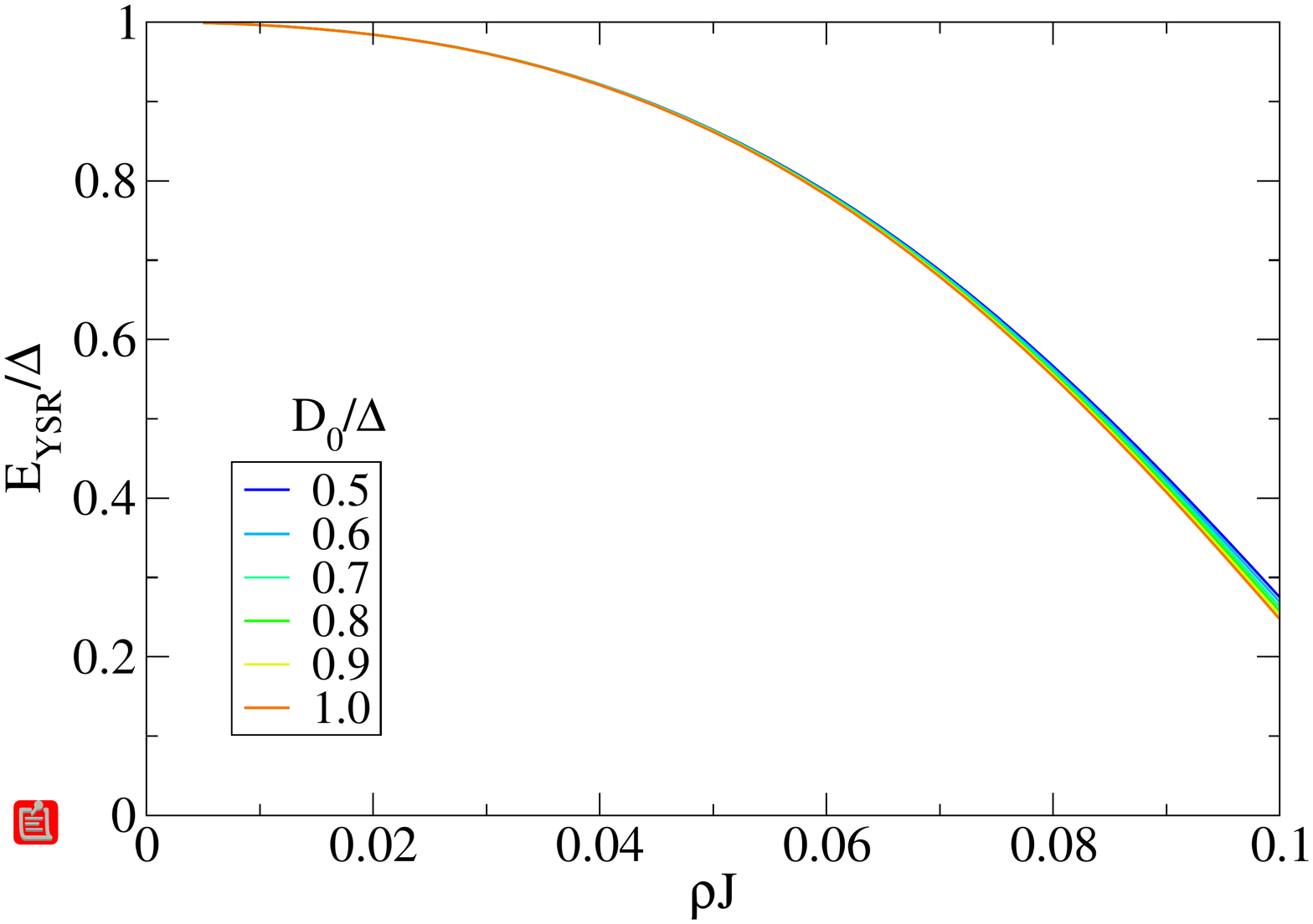}
		\caption{Yu-Shiba-Rusinov state energy as a function of exchange $J$ for $S=3/2$ model.
			The results are extracted from the highly-accurate NRG renormalization flow diagrams. The weak dependence on the value of the bare magnetic anisotropy $D_0$ enables the use of these results as a calibration of the tip offset above MnPc in terms of the exchange coupling $J$ even in the absence of prior knowledge of the precise value of (bare) anisotropy.
		}
		\label{FigSIRZ1}
	\end{figure}

	The insensitivity of the YSR state energy on $D_0$ is illustrated in Fig.~\ref{FigSIRZ1}, which enables us to calibrate the exchange coupling based on the experimental YSR energies.
	\newpage
	\newpage

	\subsection*{Bare magnetic anisotropy $D_0$ as a function of the tip-sample distance}
	
	\begin{figure}[!h]
		\centering
		\includegraphics{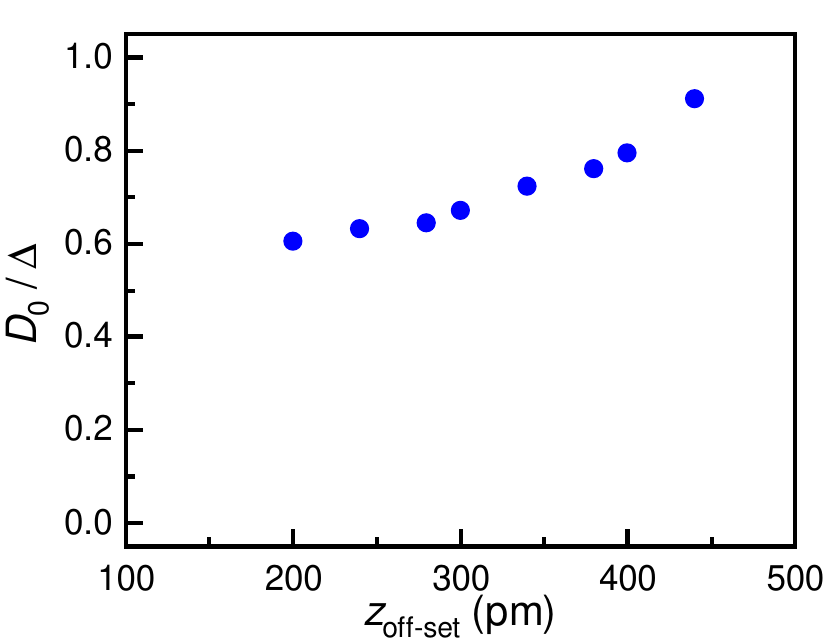}
		\caption{Bare magnetic anisotropy $D_0$. These values are extracted from the experimental $D_\mathrm{eff}/\Delta$ values by subtracting the theoretically predicted $\rho J$ dependence and plotted here as a function of the tip-sample distance offset.}
		\label{FigSID0}
	\end{figure}
	
	\newpage
	
	\subsection*{Comparison of the spectral function with normal metal and superconducting substrate}
	\begin{figure}[!h]
		\centering
		\includegraphics [width=0.6\textwidth] {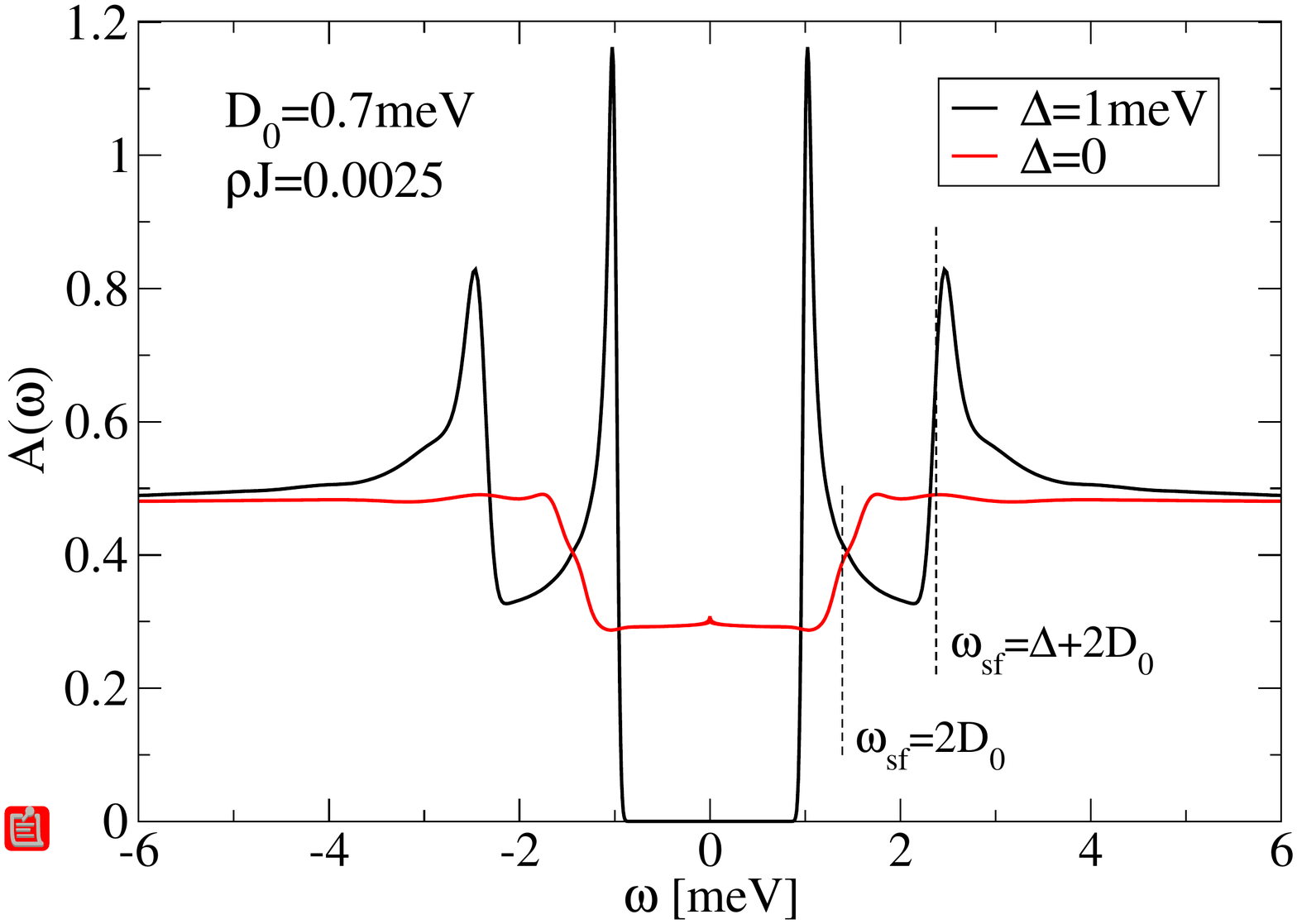}
		\caption{Calculated impurity spectral function for normal-state and
			superconducting substrate.
			We compare the low-energy parts of the spectral functions computed for
			an anisotropic Kondo impurity model ($S=3/2$, longitudinal anisotropy
			$D_0$) in the limit of very small exchange coupling $J$, so that the
			Kondo temperature is negligibly small. For normal-state substrate, the
			spectrum shows steps due to inelastic excitations for bias voltage
			beyond the spin-flip excitation threshold $\omega_\mathrm{sf}=2D_0$.
			The significant width of the step is a broadening artifact of the
			numerical method (lower broadening parameter would lead to stronger
			artifacts, which are already visible in these results in the form of
			weak oscillatory features). For superconducting substrate, a gap is
			formed, while the spin-flip features are shifted to
			$\omega_\mathrm{sf}=\Delta+2D_0$. Furthermore, the spectral shape of
			the inelastic excitations inherits the form of the density of states
			of the superconductor near the threshold of the band of Bogoliubov
			excitations.
		}
		\label{FigSIRZ2}
	\end{figure}
	
	\clearpage
	\newpage
	\subsection*{Extraction of inelastic-excitation step heights}
	\begin{figure}[!b]
		\centering
		\includegraphics [width=.9\textwidth] {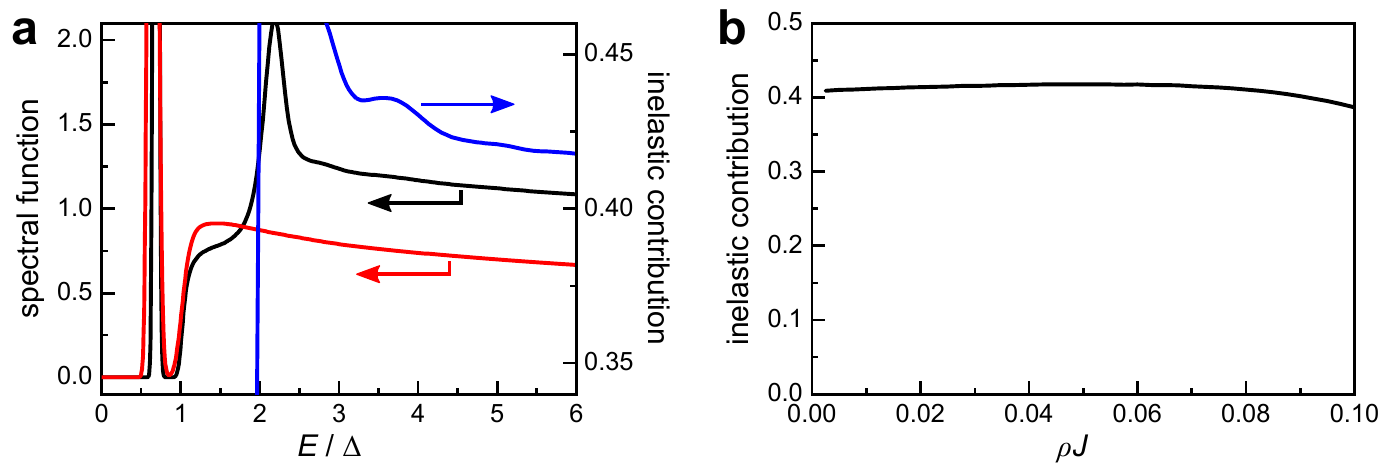}
		\caption{Estimating the real inelastic contribution to the total spectral function. (a) Calculated spectral functions with $\rho J=0.07$ and $D_0=0.7$ meV (black line) or  $D_0=5.0$ meV (red line). The difference is given by the blue line. (b) The asymptotic value of the inelastic contribution as a function $\rho J$ for $D_0=0.7$ meV.}\label{Fig_inel}
	\end{figure}
	Extracting the amplitude of the inelastic excitations in the case of a normal-state substrate is simple: extract the step amplitude and reference it to the asymptotic value. For a superconducting substrate, the inelastic excitation corresponds to a spectral feature with a complicated shape that also changes with the increasing value of the exchange coupling. 
	
	In order to compare these results with the experiments, we have used two different approaches to extract the amplitude from computed spectral functions. For low $J$ (up to $\rho J \approx 0.06$), one can define the height as the difference between the minimum value of LDOS (this point occurs between the edge of the gap and the spin-flip resonance peak) and the asymptotic value. For larger $J$, this minimum no longer exists. Instead, a plateau-like structure exists for $J$ up to $\rho J \approx 0.09$. In this range, one can define the amplitude as the difference between the value at the center of this plateau and the asymptotic value. Finally, for large $J$, no meaningful definition exists. 
	
	The maximum of the spectral function at the position of the spin excitation does not provide information about the amplitude of the spin-flip excitation, but rather reflects the properties of the bottom edge of the continuum (i.e., for small $J$, information about the superconducting coherence peak in the impurity spectral function).
	
	The procedure above yields the results shown in Fig.~4e of the main manuscript and corresponds  physically to the visibility of the inelastic features. It is in spirit similar to how the spin-flip amplitude can be (and was) estimated experimentally. On the other hand, the true inelastic contribution to the total spectral function can be assessed by comparing the calculated response to that obtained with the same $\rho J$ and large value of $D_0$. Example of such curves (at $\rho J =0.07$) is shown in Fig.~\ref{Fig_inel}a. This plot shows the calculated response with $D_0=0.7$ meV (black line) and with $D_0=5$ meV (red line). The asymptotic value of their difference (blue line) gives direct access to the inelastic contribution to the total spectral function. This is plotted as a function of $\rho J$ in Fig.~\ref{Fig_inel}b showing values very close to the expected magnitude of 0.4 for pure spin-flip excitations on a normal metal substrate.

	\subsection*{Properties of the anisotropic $S=1$ Kondo model}
	
	\begin{figure}[!h]
		\centering
		\includegraphics [width=0.5\textwidth] {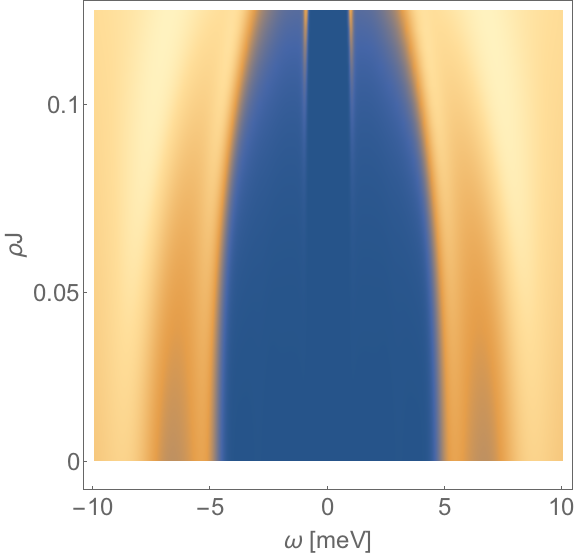}
		\caption{Theoretical spectral functions for a $S=1$ impurity with varying exchange coupling $J$. Spectral function for the anisotropic $S=1$ Kondo impurity with magnetic anisotropy parameters $D=\unit[5.5]{meV}$ and $E=\unit[1.5]{meV}$ (spin-flip excitation energies $D-E=\unit[4]{meV}$ and $D+E=\unit[7]{meV}$). At low $J$, the superconducting band edges are hardly visible, but their amplitude increases with increasing $J$. At still higher $J$ (not shown here), a $S=1/2$ sub-gap YSR state detaches from the gap edge.}
		\label{FigSIRZ6}
	\end{figure}

\bibliography{shiba}